# Speed of sound in gaseous cis-1,3,3,3-tetrafluoropropene (R1234ze(Z)) between 307 K and 420 K

D. Lozano Martín[(a)], D. Madonna Ripa[(b)], R. M. Gavioso[(b,c)]

[(a)]TERMOCAL Research Group, University of Valladolid (UVa), Paseo del Cauce 59, 47011, Valladolid, Spain

[(b)]Istituto Nazionale di Ricerca Metrologica - INRiM- Strada delle Cacce 91, 10135, Torino, Italy

[(c)]corresponding author

## Abstract

Measurements of the speed of sound in gaseous cis-1,3,3,3-tetrafluoroprop-1-ene, (R1234ze(Z)), are presented. The measurements were performed using a quasi-spherical acoustic resonator at temperatures between 307 K and 420 K and pressures up to 1.8 MPa. Ideal-gas heat capacities and acoustic virial coefficients over the same temperature range were directly calculated from the results. The relative accuracy of our determinations of the speed of sound $w(p,T)$ of R1234ze(Z) was approximately ± 0.02 %. The accuracy of the determination of the ideal gas heat capacity ratio $\gamma^0(T)$ was approximately ± 0.25 %. These data were found to be mostly consistent with the predictions of a fundamental equation of state of R1234ze(Z).

| **Nomenclature** | | *Greek symbols* | |
|---|---|---|---|
| $a_{x,y,z}$ | triaxial ellipsoid semi-axes | $\alpha_{th}$ | thermal expansion coefficient ($K^{-1}$) |
| $A_i$ | coefficients of acoustic virial equation | $\beta_a$ | 2nd acoustic virial coefficient ($m^3\ mol^{-1}$) |
| $C_p$ | molar isobaric heat capacity ($J\ mol^{-1}\ K^{-1}$) | $\Delta$ | frequency perturbation (Hz) |
| $C_V$ | molar isochoric heat capacity ($J\ mol^{-1}\ K^{-1}$) | $\varepsilon_{1,2}$ | geometrical parameters |
| $f$ | resonance frequency (Hz) | $\gamma_a$ | 3rd acoustic virial coefficient ($m^6\ mol^{-2}$) |
| $g$ | resonance halfwidth (Hz) | $\gamma$ | ideal gas heat capacity ratio |
| $k_T$ | isothermal compressibility ($Pa^{-1}$) | $\nu_i$ | coefficients of an exponential ideal gas heat capacity interpolating equation |
| $M$ | molar mass ($kg\ mol^{-1}$) | *Subscripts* | |
| $M_i$ | coefficients of 2nd acoustic virial interpolating equation | c | combined |
| $N$ | radial mode index | calc | determined from theory |
| $N_i$ | coefficients of a quadratic ideal gas heat capacity interpolating equation | EoS | equation of state |
| $p$ | pressure (Pa) | exp | determined from experiment |
| $R$ | molar gas constant ($J\ mol^{-1}\ K^{-1}$) | fit | determined from a regression |
| $T$ | temperature (K) | r | relative |
| $u_r$ | relative combined standard uncertainty | ref | reference thermodynamic variable or state |
| $u$ | standard uncertainty | th | thermal boundary layer |
| $U$ | expanded uncertainty | sh | shell |
| $w$ | speed of sound ($m\ s^{-1}$) | *Superscripts* | |
| $z$ | acoustic mode eigenvalue | 0 | ideal gas |

## 1. Introduction

The search for environmentally sustainable refrigerants has motivated a research program at INRIM to address the accurate determination of the thermodynamic properties of several candidate replacements characterized by a low global warming potential (GWP). Within this context of activity, we have measured the vapor phase speed of sound *w* in a fluorinated isomer of propene, cis-1,3,3,3-tetrafluoroprop-1-ene, conventionally referred to as R1234ze(Z). Accurate measurements of the speed of sound and the density in the compressed liquid phase of R1234ze(Z) had previously been determined at INRiM by Lago et al. (2016) and Romeo et al. (2017). Altogether these thermodynamic data will contribute to improve and update (Akasaka and Lemmon, 2018 in print) a dedicated fundamental equation of state (EoS) for this fluid, currently available from the previous work of Akasaka et al. (2014) and based on the ensemble of the experimental thermodynamic data collected for R1234ze(Z) until 2013. By that time, speed of sound data in gaseous R1234ze(Z) had been obtained only between 278.15 K and 318.15 K, as anticipated by Kayukawa et al. (2012) and privately communicated to us by Kano (2012). Extending beyond this range, the thermodynamic region explored by our measurements of $w(p,T)$ spans along nine isotherms in the temperature interval between 307 K and 420 K for pressures as high as 1.8 MPa (see Fig. 1). This region was chosen in consideration of the possible industrial application of R1234ze(Z) in heat pumping systems working at moderately high temperature, as suggested by Brown et al. (2009) and illustrated by Fukuda et al. (2014).

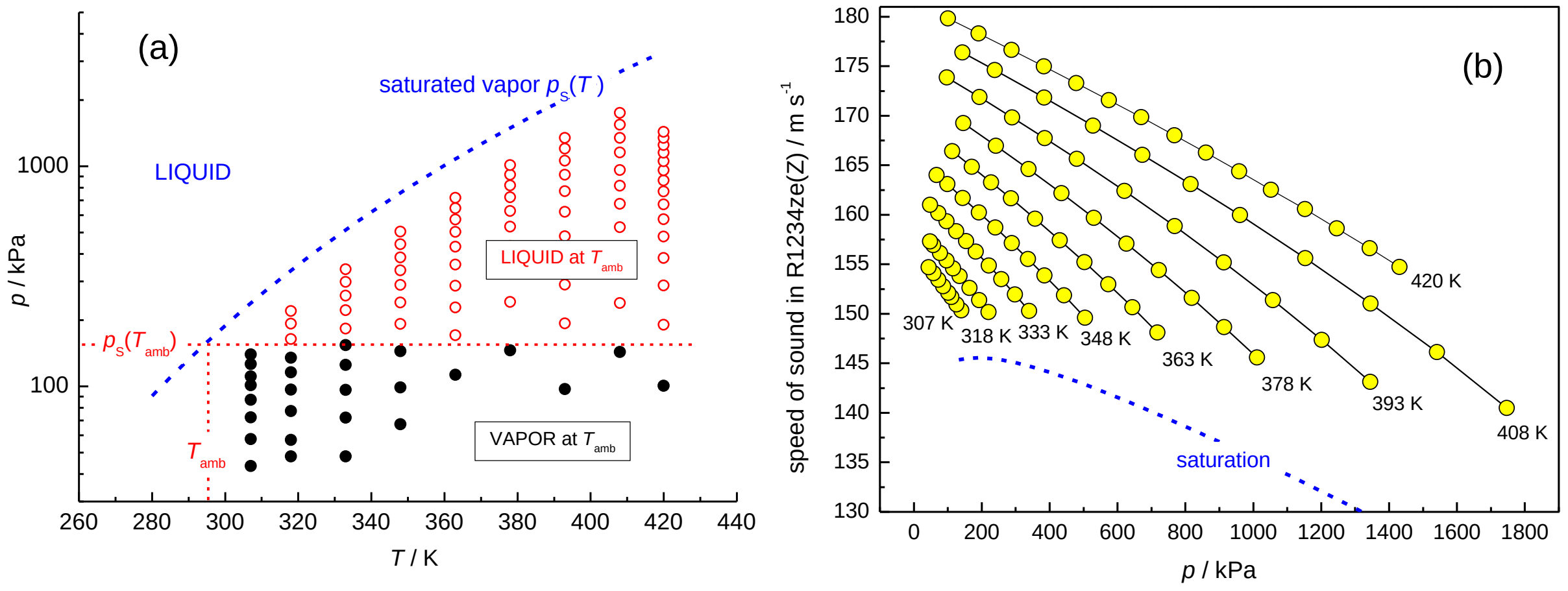

**Fig. 1** − Thermodynamic region investigated in this work. The dashed line represents the saturation curve. (a) Solid and hollow symbols respectively indicate where R1234ze(Z) is a liquid or a gas at ambient temperature; (b) Speed of sound data $w(p,T)$ for R1234ze(Z) measured along nine isotherms between 307 K and 420 K. Solid lines serve as a guide for the eye.

As for the apparatus and procedure used to determine the speed of sound of R1234ze(Z), we have chosen to measure the acoustic resonance frequencies of a gas-filled cavity with quasi-spherical shape and internal radius of 40 mm. This choice is motivated by the superior metrological performance of this method after the remarkable improvement of the physical model and the experimental procedure developed at NIST by Moldover et al. (1986). Following this achievement, which represents a milestone of physical acoustics, resonators with various internal geometries have been extensively used, proving extremely accurate, for the determination of the speed of sound of a variety of gaseous substances, including refrigerants (see e.g. Goodwin and Moldover 1990, 1991a, 1991b, Gillis 1997, Grigiante et al. 2000). With a few exceptions, the relative uncertainty[1] of our determinations of $w(p,T)$ in R1234ze(Z) is in the order of 0.02 % or less, with a dominant contribution to the uncertainty budget from our imperfect determination of the gas pressure which is discussed in the following section.

Each set of acoustic data collected at several different pressures $p$ and nearly constant temperature $T$ was analyzed to determine the ideal-gas heat capacity ratio $\gamma^0(T)= C_p{}^0(T)/C_V{}^0(T)$ and the second and third acoustic virial coefficients, $\beta_\mathrm{a}(T)$ and $\gamma_\mathrm{a}(T)$ respectively, at the same temperature, by interpolating with a polynomial function of the pressure

$$w^2\left(p,T\right)=\gamma^0 RT/M\left[1+\left(\beta_\mathrm{a}/RT\right)p+\left(\gamma_\mathrm{a}/RT\right)p^2+...\right] \quad (1)$$

where $M$ = 114.0416 × 10$^{-3}$ kg mol$^{-1}$ is the molar mass of R1234ze(Z) and $R$ = 8.3144626 J mol$^{-1}$ K$^{-1}$ is the molar gas constant (Newell et al., 2018). At all temperatures, the relative fitting precision of our determinations of $\gamma^0(T)$ resulted less than 0.03%, while the corresponding accuracy is limited by the imperfect knowledge of the composition of the sample under test, as commented below.

[1]Unless otherwise stated, all uncertainties in this work are standard uncertainties with coverage factor $k$ = 1 corresponding to a 68% confidence interval.

## 2. Experimental apparatus and measurement procedure

The basic components of the experimental apparatus, schematically drawn in Fig. 2, comprise: the acoustic/microwave resonator contained within a vacuum- and pressure-tight stainless steel vessel designed to operate up to 2.5 MPa at 450 K; a gas manifold designed to fill and evacuate the cavity by connection to a sampling bottle and a vacuum pump; a differential pressure transducer inserted in a separate container and there maintained at constant temperature near 420 K by an electrical heater. Between 310 K and 420 K, a circulating thermostatting unit precisely controls the temperature of a stirred liquid bath where the apparatus is immersed. This configuration prevented the condensation of the sample above the saturation pressure of R1234ze(Z) at ambient temperature, which is approximately 150 kPa (see Fig. 1), and facilitated the calibration of the differential pressure transducer.

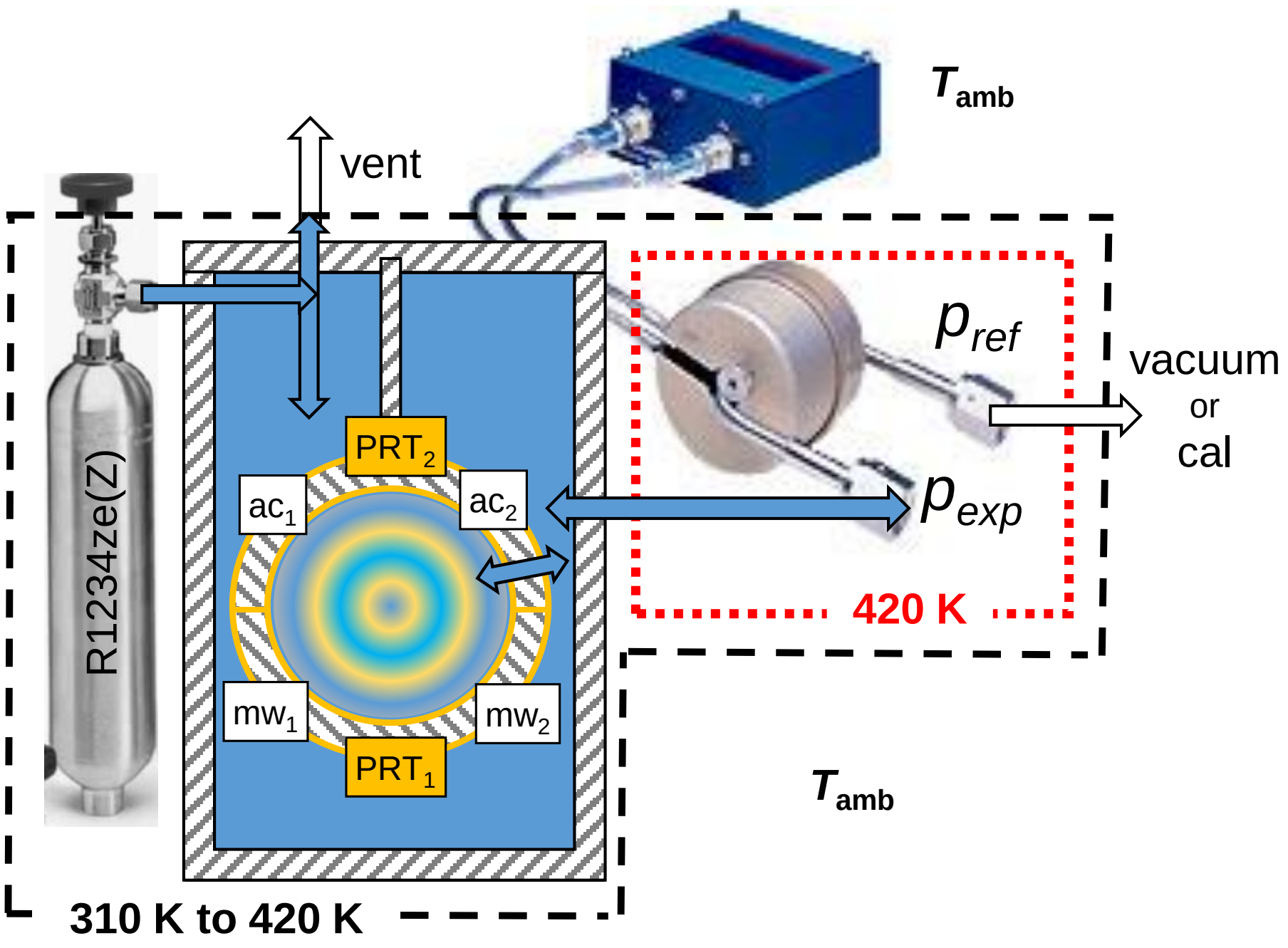

**Fig. 2** − Schematic view of the experimental apparatus. Striped patterns show the cross-section of the pressure vessel and the acoustic/microwave resonator. Text labels display the position of two 1/4" capacitance microphones (ac$_{1,2}$), two microwave loop probes (mw$_{1,2}$) and two capsule platinum resistance thermometers (PRT$_{1,2}$). To prevent condensation of gaseous R1234ze(Z), the elements within the black dashed line (sampling bottle, vessel, resonator, tubing) were maintained at the same temperature, between 310 K and 420 K, by a liquid bath thermostat. To avoid the need of multiple calibrations, the differential pressure transducer was always maintained near 420 K by a dedicated controlled heater.

### 2.1 Resonant cavity

The resonator used in this work is a hollow cavity assembled from two nearly hemispherical pieces constructed in 316L-type stainless steel. The internal shape of the cavity, with a volume of approximately 268 $cm^3$, is designed as a triaxial ellipsoid with semi-axes nominally equal to $a_x$ = 40.00 mm, $a_y = a_x(1+\varepsilon_1)$ = 40.06 mm and $a_z = a_x(1+\varepsilon_2)$ = 40.12 mm. This type of geometry, first suggested for gas metrology by May et al. (2004), improves the precision achievable in the determination of microwave resonances by lifting their intrinsic degeneracy. The cavity surface was gold-plated, internally and externally, with a layer approximately 10 μm thick, to enhance the quality factors of the microwave resonances and increase corrosion resistance and chemical inertness. Four ports on the cavity wall and suitable adapters accommodated two condenser microphones and two antennas, as needed to excite and detect the acoustic and microwave field. Three additional ducts, with a diameter of 1.5 mm, provided access to the cavity interior for gas filling and evacuating. To minimize their perturbation onto the acoustic resonances, the length of two of these ducts was extended to approximately equal the internal radius of the cavity using short sections of 1/8" o. d. stainless steel tubes. A longer section of 1/8" PTFE tube was connected to the third duct to flow the gas under test directly from the external manifold into the cavity. A set of electrical feed-throughs, welded to the top plate of the pressure vessel, provided a gas-tight connection to the transducers within the cavity and drive the signals, across the thermostatted bath, to the instrumentation in the laboratory.

### 2.2 Temperature and pressure measurement and control

The temperature of the liquid within the bath (a mixture of water and ethylene glycol between 307 K and 348 K; Dow Corning Xiameter PMX 200[2] silicon oil between 363 K and 420 K), was controlled by a Julabo SE-Z heating circulator with a set-point precision of ± 0.01°C.

The temperature of the gas within the resonator was inferred from the resistance readings of two 100 Ω capsule-type platinum resistance thermometers (PRT) inserted into cylindrical

[2]Identification of commercial equipment and materials does not imply recommendation or endorsement by INRiM nor does it imply that the equipment and materials identified are necessarily the best available for the purpose

extensions of the top and bottom ends of the resonator, and previously calibrated at INRiM by comparison to a secondary national standard. The resistance of the PRTs was read by a Keithley 2700 multimeter, whose calibration in the 100 Ω range was checked by comparison with a Tinsley 5685 standard reference resistor. The stability of the resonator temperature, over the short time - less than 1 min - required to record and fit the resonance data of each acoustic or microwave mode, was typically well below 1 mK, allowing an extremely precise determination of the resonance parameters.

At all pressures and temperatures explored in this work, the temperature difference between the two PRTs was found in the order of a few mK and, in any case, less than the combined calibration uncertainty of the thermometers.

The pressure of the gas was measured using a differential gauge (MKS 616A) based on a capacitance diaphragm sensor with a full scale range of 1.98 MPa. This gauge can be continuously operated up to 300 °C, thanks to the separation of the physical sensor from the electronic circuitry by triaxial cables, avoiding condensation of the sample under test. In operation, the measurement side of the gauge was kept in contact with the gas by a tube immersed and thermostatted within the bath, while the reference side of was continuously maintained below 5 Pa by a dry scroll pump.

The manufacturer specification of the relative accuracy of the capacitance gauge is 0.12% of the reading. However, previous calibrations and the experimental practice showed the response of this gauge to be significantly temperature-dependent. To improve beyond this limited performance, we isolated the pressure gauge from direct contact with the liquid bath using a metal container continuously thermostatted at 420 K by a dedicated heater. In the same condition, we calibrated the capacitance gauge by comparison with a more accurate transducer, a Paroscientific Digiquartz 745-300A with a full scale range of 2.1 MPa. The calibration results were used to determine a linear correction factor (1.0402 ± 0.0001), with the corresponding contribution to the uncertainty budget discussed below.

### 2.3 Sample purity

The test sample of R1234ze(Z) was provided by Central Glass Co. Ltd. with a declared purity greater than 0.99 in mass fraction and was taken from the same batch used to measure speed of sound and density in the liquid phase (Lago et al., 2016; Romeo et al., 2017) and vapor pressure (Fedele et al., 2014). No further analysis was conducted but, upon transfer of approximately 1 kg of sample into a stainless steel bottle, repeated freeze-pump-thaw cycles by immersion in a dewar filled with liquid $N_2$ were found effective in removing about 13 g of volatile impurities, a quantity which is commensurate with the estimated (fractional) amount of residual impurities declared by the manufacturer. This datum, and the good agreement of the present measurements with the prediction of the EoS of Akasaka et al. (2014), constructed from a variety of experimental results, including some which are rather insensitive to impurities, e. g. speed of sound in the compressed liquid, is reassuring. The impact of possible residual non-volatile impurities upon our speed of sound and heat capacity measurements is impossible to estimate, in absence of additional information about the identity and the amount of each impurity. In the following, we assume that a plausible estimate of the relative uncertainty of the molar mass $u_r(M)$ is 100 ppm. Such a relative variation of $M$ would for instance be caused by a neglected impurity having a concentration of 0.02 %, assuming that its molar mass $M_{imp}$ differs from $M$ by 50 %.

### 2.4 Microwave determination of the resonator dimensions

Two microwave antennas, bent in the form of a loop, protruded by a few mm within the cavity and suitably coupled to both TM and TE modes. Standard instrumentation and procedures, as in May et al. (2004), were used to acquire microwave data and determine the resonance parameters of four triply-degenerate modes TM11, TM12, TE11 and TE12 at frequencies between 3.3 GHz and 9.2 GHz. By measuring the microwave resonance frequencies of the cavity, the internal radius $a$ can be determined, as needed to obtain speed of sound estimates from the measured acoustic frequencies. The data displayed in Fig. 3 show the variation of the internal cavity radius between 307 K and 420 K, with each datum obtained by averaging the results of four modes, after applying corrections to account for the relevant perturbations, including the finite electrical conductivity of the surface, the quasi-spherical geometry and the admittance of ducts and waveguides. Remarkably, the entire data set can be adequately fit with a simple quadratic function of the temperature to provide an estimate of the radius of the cavity with relative precision in the order of 1 part in $10^6$. As an example, the radius of the cavity at 360 K and zero pressure is estimated: $a(0, 360\text{ K})$ = (40.091 567 ± 0.000 045) mm. The linear term of the interpolation $\alpha_{th}$ = (16.89 ± 0.02)·$10^{-6}$ $K^{-1}$ estimates very precisely the thermal expansion of the assembled cavity, in good agreement with tabulated values (Desai and Ho, 1978) of the coefficient of thermal linear expansion of 316L

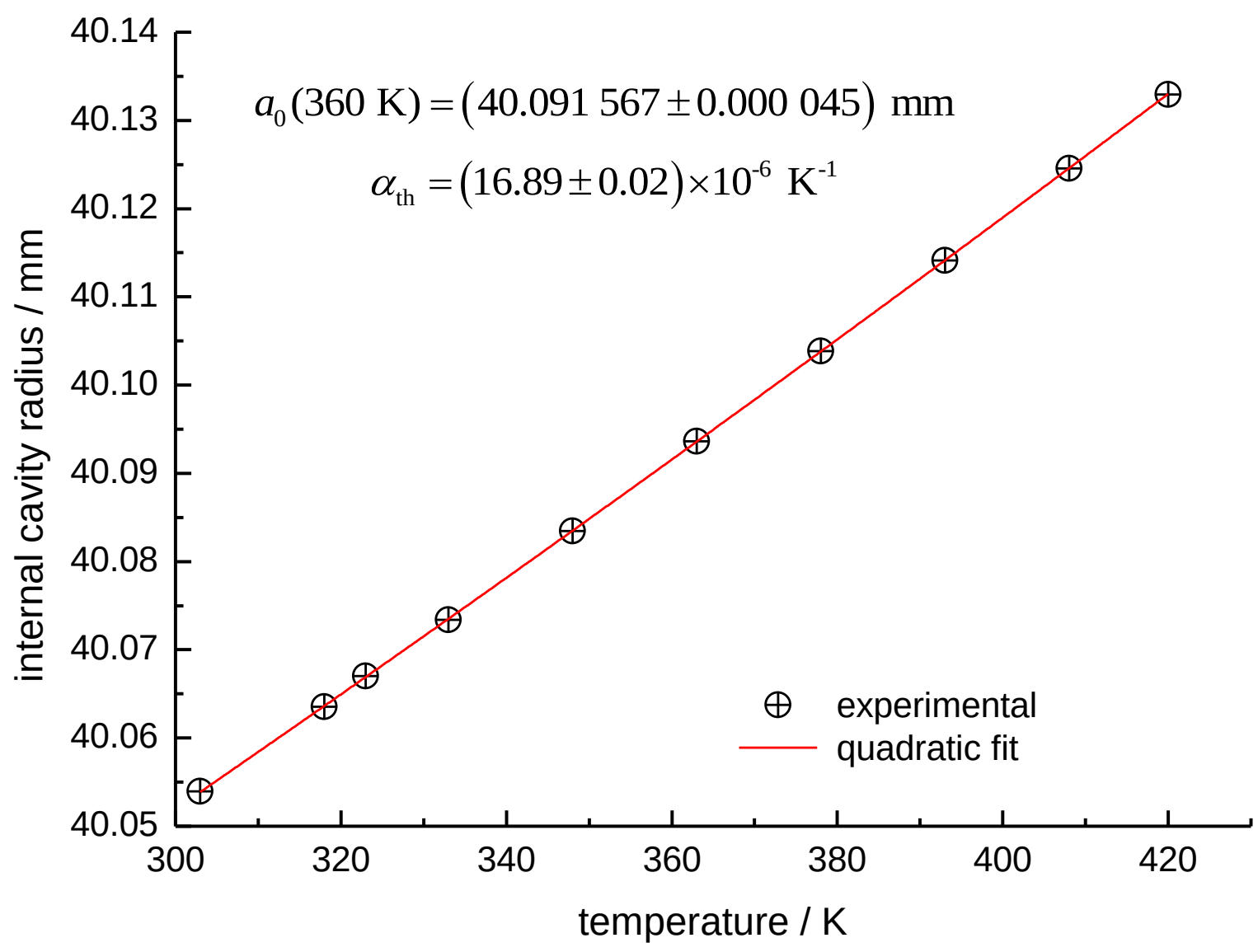

stainless steel in this temperature range.

**Fig. 3** − Microwave determination of the internal cavity radius. At each temperature the crossed circle symbols shows the mean radius determined from several microwave modes. The solid line displays the result of a quadratic fit as a function of temperature. Text labels report the estimates of the radius of the evacuated cavity $a_0$ at 360 K, and the linear coefficient of thermal expansion $\alpha_{th}$.

The internal radius of the cavity decreases as a function of increasing pressure, as the gas fills both the interior of the cavity and the outer vessel. The volume compressibility of 316L steel measured by Ledbetter (1981) near ambient temperature $k_T$ = 6.3·$10^{-12}$ $Pa^{-1}$ indicates that at 1.8 MPa, the highest pressure explored in this work, the corresponding relatively variation of the radius is only 2 ppm and was included for completeness. Altogether, the dimensional parameters reported and discussed above were combined to predict the variation of the internal radius of the cavity as a function of temperature and pressure:

$$a(p,T) = a_0(T_{\text{ref}})\left[1+\alpha_{\text{th}}(T-T_{\text{ref}})\right]\left(1+\frac{k_T}{3}p\right)^{-1} \quad (2)$$

Finally, the frequency separation between the single peaks within each microwave triplet were analyzed to determine the internal shape of the cavity, resulting in the mean geometrical parameters $\varepsilon_1 = 2.4\cdot10^{-3}$, $\varepsilon_2 = 6\cdot10^{-4}$, which differ from the designed specification $\varepsilon_1 = 3.0\cdot10^{-3}$, $\varepsilon_2 = 1.5\cdot10^{-3}$ requested to the mechanical shop. The estimated values of $\varepsilon_1$, $\varepsilon_2$ were used to calculate the acoustic eigenvalues $z_{0N}$ of each radial mode (0, $N$) using the perturbation theory worked out by Mehl, (2007). The relative differences of $z_{0N}$ from the corresponding eigenvalues of a perfect sphere vary between 2 ppm for the (0,2) mode and 87 ppm for mode (0,10).

**2.5 Acoustic model**

Two condenser microphones, mounted with their diaphragm flush with the internal cavity surface, were used as acoustic transducers. The source was a 1/4" free-field cartridge (Brüel & Kjaer 4939), kept unpolarized to vibrate at twice the frequency of the electrical signal fed from an external amplifier, reducing the cross-talk to the detector microphone, initially a pre-polarized 1/4" free-field (GRAS 40BE) which, due to malfunctioning, had to be replaced above 348 K with a 4939 cartridge externally polarized with 200 DCV.

Standard instrumentation and procedures (see e. g. Moldover et al., 2014) were used for the acquisition and the interpolation of the acoustic resonance frequencies $f_{0N}$ and halfwidths $g_{0N}$ of up to nine purely radial (0,$N$) modes at frequencies comprised between a minimum of 2.7 kHz for the (0,2) mode and 21.3 kHz for mode (0,10). At all temperatures and pressures explored in this work the relative fitting precision of a single acoustic resonance frequency of any mode was always significantly less than 1 part in $10^6$ and as such, negligible with respect to other contributions to the acoustic uncertainty budget. However, at any ($p$, $T$) state, the relative standard deviation of several repeated acquisitions of the resonance frequency of any mode was found to vary between a few ppm and a few parts in $10^5$, reflecting our limited capability to maintain constant and precisely reproduce the gas pressure, which is kept into account with a specific uncertainty contribution discussed below. Thus, upon correcting each repeated frequency measurement to the same reference temperature $T_{\mathrm{ref}}$ chosen for the isotherm (see Table 1) and averaging the corrected data set to a single mean estimate of $f_{0N}$ and $g_{0N}$ for every mode at every ($p$, $T$) state, a corresponding speed of sound value $w_{0N}$ was calculated for each mode as

$$w_{0N} = 2\pi a\left( f_{0N} - \Delta f_{th} - \Delta f_{sh} \right) / z_{0N} \tag{3}$$

where $\Delta f_{\rm th}$ and $\Delta f_{\rm sh}$ are small correction terms which account for the frequency perturbation respectively induced by: heat exchange between the gas and the shell in a narrow boundary layer adjacent to the cavity wall; coupling of acoustic resonances in the cavity with elastic resonances within the shell.

In Eq. (3), the cavity radius $a = a(p,T)$ and the acoustic eigenvalues $z_{0N}$ were determined by microwave measurements, as discussed above.

The evaluation of the thermal boundary layer correction $\Delta f_{\rm th}$ required: i) an estimate of thermal conductivity of dilute R1234ze(Z), which was first available in 2015 when the Reference database for the Thermodynamic and Transport properties of Fluids (REFPROP) by Lemmon et al. (2013) was updated to include the fluid R1234ze(Z); ii) an estimate of the density and heat capacity of gaseous R1234ze(Z), which is available from the dedicated EoS of Akasaka et al., 2014; iii) an estimate of smaller contributions due to: the temperature-jump effect, accounted by setting the thermal accommodation coefficient equal to unity; the propagation of a thermal wave into the shell, whose account requires tabulated values of the heat capacity and thermal conductivity of gold (Kaye and Laby, 1995).

The evaluation of the frequency perturbations induced by the coupling of gas and shell motion $\Delta f_{\rm th}$ required the estimate $\chi_{\rm i} = 3.0\cdot10^{-11}$ Pa$^{-1}$ of the elastic compliance of the shell under the action of the internal (acoustic) pressure, as obtained from an elastic model and published values of the properties of 316L steel (see e.g. Moldover et al, 1986 and Ledbetter et al., 1975), and the calculation of the breathing frequency of the shell $f_{\rm br}$ = 30.4 kHz, based on the model of Mehl (1985). Given the rather low speed of sound of R1342ze(Z), the frequency spectrum of all the radial modes examined in this work falls much lower than $f_{\rm br}$ and the magnitude of the shell perturbations $\Delta f_{\rm th}/ f_{0N}$ is relatively small, resulting at maximum 24 ppm for the highest frequency mode (0,10) at 420 K.

Finally, at one intermediate test isotherm 363 K, we estimated the frequency perturbations induced by the finite acoustic impedance of three gas-inlet ducts and two microphone diaphragms using the methods and the models respectively described by Gillis et al. (2009) and Guianvarc'h et al. (2009) and found them to be less than 10 ppm, with correspondingly small contributions to the resonance halfwidths, and as such, safely negligible in the present uncertainty context.

The adequateness and completeness of the acoustic model described above can be assessed by two relevant indicators, namely the dispersion of the speed of sound $w_{0N}$ determined from nine radial modes around their mean $\langle w \rangle = \sum_N w_{0N} / 9$ and the difference between the experimental and the calculated resonance halfwidths

$$\Delta g = g_{\mathrm{exp}} - g_{\mathrm{calc}} = g_{\mathrm{exp}} - \left( g_{\mathrm{th}} + g_{\mathrm{bulk}} \right), \tag{4}$$

where $g_{\mathrm{th}} = -\Delta f_{\mathrm{th}}$ accounts for heat losses in the boundary layer, and $g_{\mathrm{bulk}}$ accounts for viscous and thermal energy losses in the bulk of the fluid.

At the lowest and highest temperature explored in this work, 307 K and 420 K, the relative speed of sound dispersion and the relative excess halfwidths are respectively displayed in Fig. 4 and Fig. 5. It is apparent that, in all instances, the speed of sound estimated from the acoustic data of mode (0,2) significantly differ from that obtained from the other modes and have associated correspondingly larger excess halfwidths. These evident discrepancies, caused by a combination of unknown imperfections of our model which appear particularly relevant in the low frequency range, motivated the rejection of mode (0,2) data from any following computation.

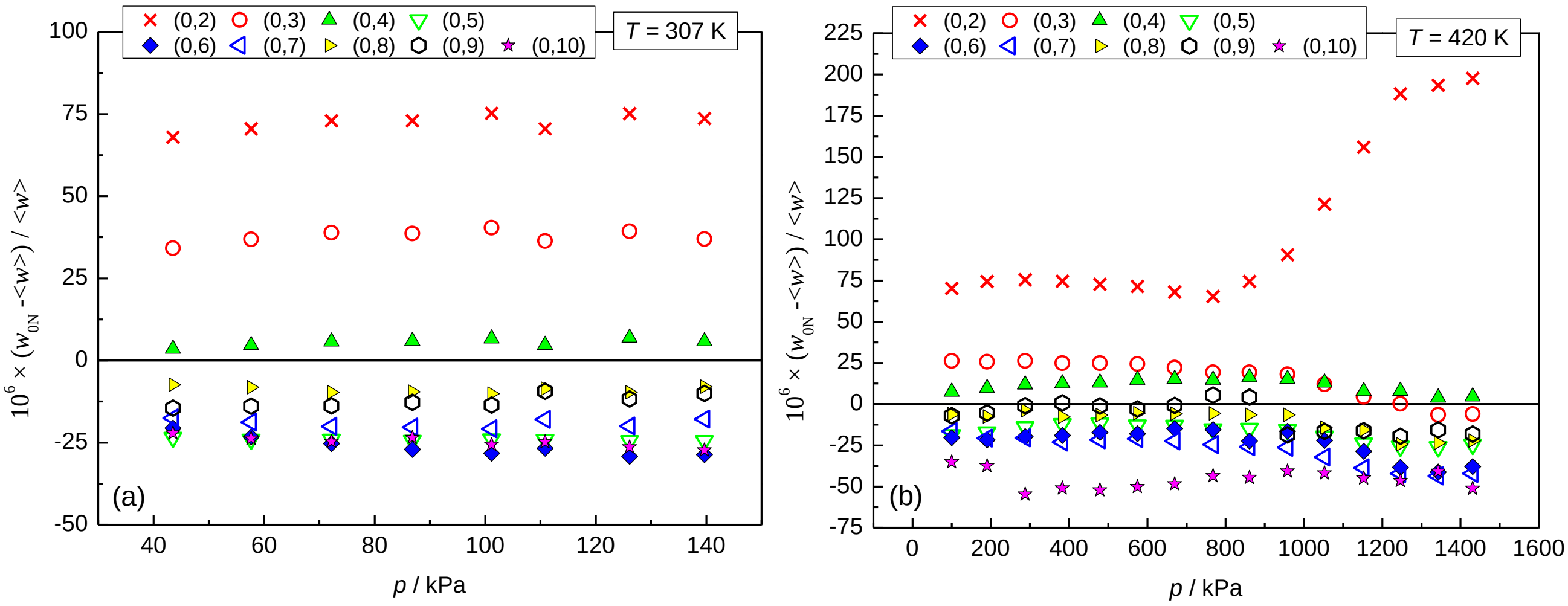

**Fig. 4**. − Relative dispersion of speed of sound $w_{0N}$ estimated from nine radial acoustic modes with respect to their mean value $<w>$ at: (a) $T = 307$ K; (b) $T = 420$ K.

For the remaining modes, the relative dispersion of the speed of sound data is in the order of ± 30 ppm and the corresponding excess-halfwidths is always less than 40 ppm, with the magnitude of both indicators rather constant as a function of temperature and pressure. These typical figures set a lower limit to the accuracy of the present measurements and discouraged from trying to deduce information on vibrational relaxation using the frequency dependence of the excess halfwidths. This additional approximation of our model has no practical consequences on the measured speed of sound values or their associated uncertainty.

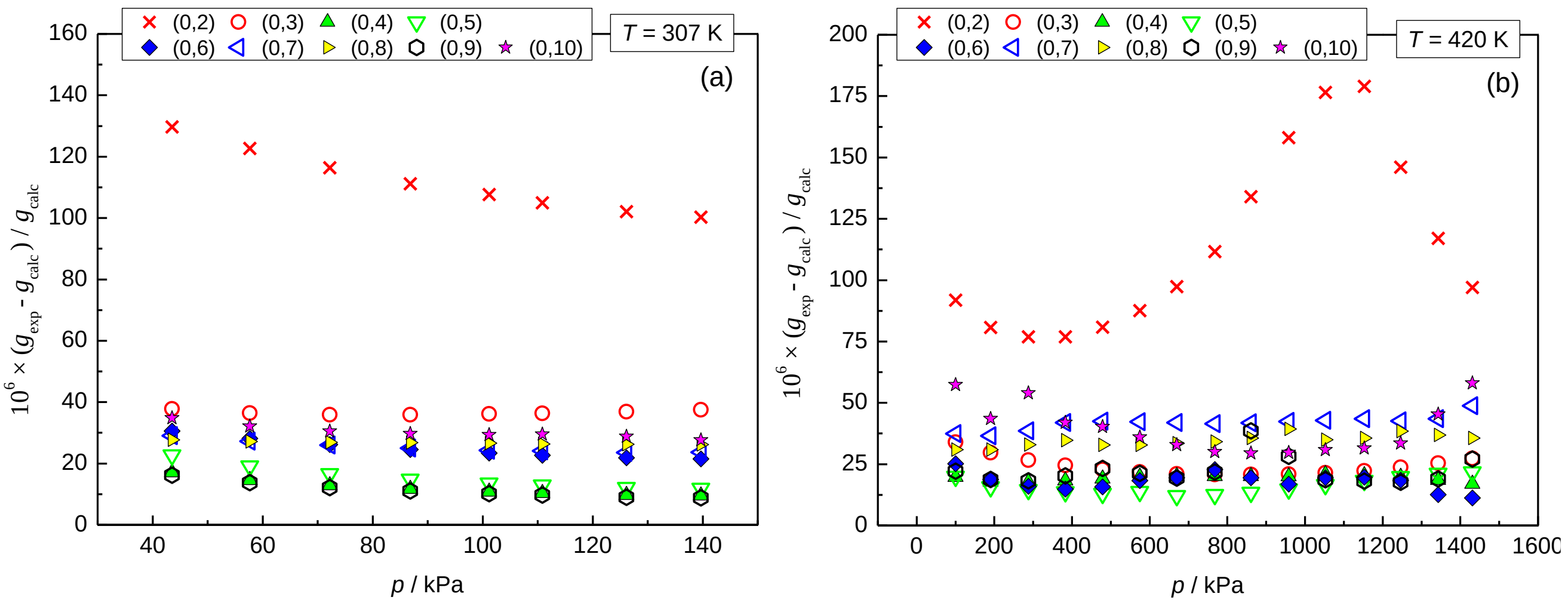

**Fig. 5**. − Relative excess halfwidths $(\Delta g/f)_{0N}$ of nine radial acoustic modes at: (a) $T$ = 307 K; (b) $T$ = 420 K.

## 3. Results

Before each series of speed of sound measurements at different pressures along an isotherm, the resonator was thoroughly evacuated by a rotary vane pump, below 1 Pa, for at least 12 hours, while microwave data and the zero indication of the capacitance pressure gauge were recorded. Subsequently, the resonator was filled up to the highest measurement pressure for the isotherm, set to approximately the 70 % of the estimated saturation pressure of R1234ze(Z), to avoid pre-condensation (see Mehl and Moldover, 1982). Following the initial compression of the fluid, and after each expansion to a lower pressure step along the isotherm, a few hours were needed for the resonator to re-equilibrate with the bath to a stable temperature condition. When this condition was met, typically after two hours, the recording of temperature, pressure, and acoustic resonance data was started and continued for at least three hours, providing redundant measurement data which allowed to estimate the signal/noise ratio, repeatability and possible drifts. Each isotherm was concluded getting back to vacuum to evaluate the possible change of the zero indication of the pressure-gauge.

### 3.1 Speed of sound

For nine isotherms between 307 K and 420 K and a variable number of pressures for each isotherm, the speed of sound of R1234ze(Z) was estimated from the acoustic frequencies of nine radial modes using Eq. (3). Upon rejection of mode (0,2) data, an average estimate $w(p, T)$ was calculated as the arithmetic mean of the data of modes (0,3) to (0,10). These estimates and their relative combined standard uncertainties $u_{c,r}(w_{exp})$ are listed in Table 1.

Looking for a single aggregate indicator of the uncertainty of the speed of sound of R1234ze(Z) determined in this work, we considered the mean relative expanded ($k$ = 2) uncertainty $U_{c,r}$ = 0.037 % obtained by averaging the relative expanded uncertainty of all the speed of sound dataset.

Table 1 – Speed of sound $w_{exp}$ of R1234ze(Z), relative combined standard uncertainty $u_{c,r}(w_{exp})$ and relative differences $(w_{exp} - w_{EoS})/w_{EoS} = \Delta w/w_{EoS}$ from the speed of sound predicted by the EoS of Akasaka et al., 2014. The listed relative combined uncertainties $u_{c,r}(w_{exp})$ result from the quadrature sum of several contributions including the imperfect determination of temperature (typically ±20 mK), pressure (up to ±1.5 kPa), molar mass (100 ppm), and acoustic frequencies (always < 50 ppm).

| $p$<br>kPa | $w_{exp}$<br>m s$^{-1}$ | $u_r(w_{exp})$<br>ppm | $\Delta w/w_{EoS}$<br>ppm | $p$<br>kPa | $w_{exp}$<br>m s$^{-1}$ | $u_c(w_{exp})$<br>ppm | $\Delta w/w_{EoS}$<br>ppm |
|---|---|---|---|---|---|---|---|
| $T$ = 307.00 K | | | | $T$ = 378.00 K | | | |
| 43.52 | 154.701 | 222 | −299 | 145.70 | 169.259 | 124 | 18 |
| 57.67 | 154.083 | 225 | −308 | 241.83 | 166.973 | 128 | −47 |
| 72.22 | 153.438 | 230 | −324 | 337.95 | 164.624 | 133 | −108 |
| 86.82 | 152.779 | 233 | −368 | 434.88 | 162.179 | 139 | −191 |
| 101.20 | 152.115 | 239 | −451 | 530.58 | 159.674 | 151 | −351 |
| 110.86 | 151.682 | 240 | −388 | 626.46 | 157.084 | 189 | −470 |
| 126.16 | 150.956 | 245 | −499 | 721.98 | 154.417 | 211 | −533 |
| 139.66 | 150.348 | 252 | −323 | 818.53 | 151.593 | 174 | −710 |
| $T$ = 318.00 K | | | | 914.54 | 148.663 | 198 | −872 |
| 48.12 | 157.291 | 73 | −287 | 1011.10 | 145.573 | 201 | −1024 |
| 57.09 | 156.936 | 87 | −320 | $T$ = 393.00 K | | | |
| 77.22 | 156.137 | 77 | −361 | 97.15 | 173.877 | 118 | 257 |
| 96.48 | 155.372 | 86 | −343 | 193.23 | 171.877 | 113 | 147 |
| 115.75 | 154.582 | 86 | −407 | 289.46 | 169.830 | 118 | 44 |

| | | | |
|---|---|---|---|
| 134.76 | 153.804 | 84 | −394 |
| 163.84 | 152.595 | 83 | −352 |
| 192.46 | 151.373 | 85 | −328 |
| 219.83 | 150.175 | 96 | −304 |
| *T* = 333.00 K | | | |
| 48.02 | 160.991 | 172 | −257 |
| 72.13 | 160.168 | 174 | −307 |
| 96.43 | 159.341 | 182 | −277 |
| 125.31 | 158.326 | 182 | −357 |
| 153.98 | 157.322 | 189 | −306 |
| 182.92 | 156.280 | 190 | −327 |
| 221.47 | 154.871 | 197 | −308 |
| 257.94 | 153.493 | 201 | −364 |
| 298.30 | 151.935 | 210 | −379 |
| 340.13 | 150.272 | 227 | −393 |
| *T* = 348.00 K | | | |
| 67.35 | 164.024 | 150 | −144 |
| 98.90 | 163.083 | 154 | −166 |
| 144.47 | 161.701 | 161 | −201 |
| 192.18 | 160.220 | 165 | −258 |
| 240.41 | 158.696 | 168 | −281 |
| 288.78 | 157.128 | 173 | −322 |
| 336.60 | 155.534 | 182 | −395 |
| 385.06 | 153.881 | 190 | −444 |
| 442.38 | 151.855 | 205 | −557 |
| 504.40 | 149.603 | 221 | −537 |
| *T* = 363.00 K | | | |
| 112.94 | 166.407 | 248 | −51 |
| 170.56 | 164.854 | 256 | −189 |
| 228.13 | 163.276 | 265 | −300 |
| 286.25 | 161.647 | 274 | −427 |
| 357.71 | 159.584 | 283 | −645 |
| 430.21 | 157.430 | 298 | −862 |
| 502.61 | 155.223 | 313 | −1001 |
| 573.08 | 152.981 | 323 | −1268 |
| 644.37 | 150.649 | 349 | −1428 |
| 717.96 | 148.100 | 374 | −1899 |
| 386.09 | 167.736 | 128 | −14 |
| 479.93 | 165.638 | 150 | −160 |
| 620.79 | 162.393 | 506 | −365 |
| 768.92 | 158.845 | 347 | −520 |
| 913.37 | 155.198 | 214 | −851 |
| 1057.88 | 151.388 | 252 | −1028 |
| 1201.85 | 147.364 | 548 | −1300 |
| 1344.57 | 143.132 | 220 | −1497 |
| *T* = 408.00 K | | | |
| 143.41 | 176.374 | 66 | 128 |
| 238.68 | 174.602 | 99 | 58 |
| 384.52 | 171.818 | 74 | −103 |
| 527.93 | 169.001 | 104 | −270 |
| 673.72 | 166.057 | 119 | −387 |
| 815.75 | 163.093 | 174 | −523 |
| 960.84 | 159.956 | 97 | −685 |
| 1153.70 | 155.606 | 116 | −866 |
| 1345.42 | 151.032 | 201 | −1090 |
| 1541.10 | 146.121 | 186 | −936 |
| 1746.72 | 140.483 | 144 | −1311 |
| *T* = 420.00 K | | | |
| 100.49 | 179.824 | 99 | 294 |
| 190.76 | 178.301 | 100 | 210 |
| 287.57 | 176.647 | 111 | 144 |
| 382.90 | 174.994 | 112 | 79 |
| 478.75 | 173.304 | 111 | 7 |
| 574.78 | 171.585 | 114 | −51 |
| 669.91 | 169.852 | 120 | −124 |
| 767.85 | 168.032 | 138 | −217 |
| 860.67 | 166.263 | 135 | −395 |
| 958.18 | 164.392 | 139 | −443 |
| 1052.50 | 162.532 | 156 | −583 |
| 1152.30 | 160.561 | 137 | −505 |
| 1246.21 | 158.620 | 279 | −723 |
| 1342.77 | 156.615 | 214 | −734 |
| 1430.73 | 154.735 | 165 | −831 |

The uncertainties in Table 1 are the quadrature sum of several contributions, assumed to be uncorrelated, including our imperfect determination of temperature and pressure, molar mass, mode-dependent inconsistencies, and the variance of repeated measurements.

The main contribution to the combined standard uncertainty of our temperature measurements comes from the calibration of the PRTs on the International Temperature Scale of 1990 (ITS-90), estimated to be 20 mK, constant over the whole temperature range explored here. A second contribution comes from the small (<5 mK) temperature gradient measured across the apparatus, estimated as the average difference, assuming a rectangular probability distribution, of the readings of the two PRTs along one isotherm. Given that $\partial w/\partial T$ varies between a minimum 0.20 m $s^{-1}$ $K^{-1}$ and a maximum of 0.45 m $s^{-1}$ $K^{-1}$, the typical combined temperature uncertainty of 20 mK contributes to the relative combined speed of sound uncertainty $u_{c,r}(w)$ with less than 100 ppm in the worst case.

The uncertainty of our pressure measurements is estimated as the quadratic sum of three sources: the first is a possible undetected systematic drift of the readings from the pressure transducer, evaluated by checking twice the zero reading of the differential pressure transducer, using vacuum as a reference, once before starting and a second time after completing measurements along each isotherm. Depending on the isotherm, the observed drifts varied between a minimum of 0.1 kPa at 318 K to a maximum of 1.5 kPa at 363 K; the second source results from the uncertainty of the linear correction determined by the calibration of the pressure gauge, with contributions varying between 0.005 kPa and 0.5 kPa; the last source is the standard deviation of several repeated pressure measurements with an average contribution of 0.25 kPa. Among these three sources, the uncertainty due to the zero drift represents the dominant contribution, with a few exceptions caused by a poor performance of the pressure control system.

With contributions to $u_{c,r}(w)$ up to 500 ppm, the uncertainty due to pressure measurements dominates the overall uncertainty budget.

Our estimate of a plausible uncertainty of the molar mass, $u_r(M)$ = 100 ppm results in a contribution of 50 ppm to all the speed of sound measurements in this work.

For each $w(p,T)$ determination we considered two additional contributions related to frequency measurements: i) mode inconsistency, exemplified in Fig. 4, which was assessed assuming a rectangular probability distribution bounded by the larger and lower speed measured by any of the modes examined; ii) the standard deviation of repeated frequency measurements, upon applying corrections to compensate for slight temperature differences. With few exceptions, both contributions are less than 0.01 m/s, or relatively 50 ppm.

The fractional deviations of the experimental speed of sound $w_{\mathrm{exp}}$ of R1234ze(Z) determined in this work from the $w_{\mathrm{EoS}}$ predicted by the fundamental Helmholtz energy EoS of Akasaka et al. (2014) are listed in Table 1 and displayed in Fig. 6. With the exception of a few states at 393 K, 408 K and 420 K for pressures lower than 300 kPa, most of the data along all the experimental isotherms show negative differences ($w_{\mathrm{exp}} - w_{\mathrm{EoS}}$) typically increasing with increasing pressure. For the four lower isotherms, between 307 K and 348 K, these deviations are found to be consistent with the model EoS, whose estimated uncertainty in representing gaseous speed of sound is 0.05 %. For the five higher isotherms between 363 K and 420 K, the consistency between the EoS and our experiment is limited to the lower pressure range, while relative deviations up to −0.18 % are found at higher pressures. However, the root mean square of the relative deviations $\Delta_{\mathrm{RMS}} = \left[ N^{-1} \sum_{i=1}^{N} \left( \left( w_{\mathrm{exp}} - w_{\mathrm{EoS}} \right) / w_{\mathrm{EoS}} \right)_i^2 \right]^{1/2}$ between the experiment and the model is $\Delta_{\mathrm{RMS}} = 0.058$ %, comparable with the 0.05% uncertainty stated by the model.

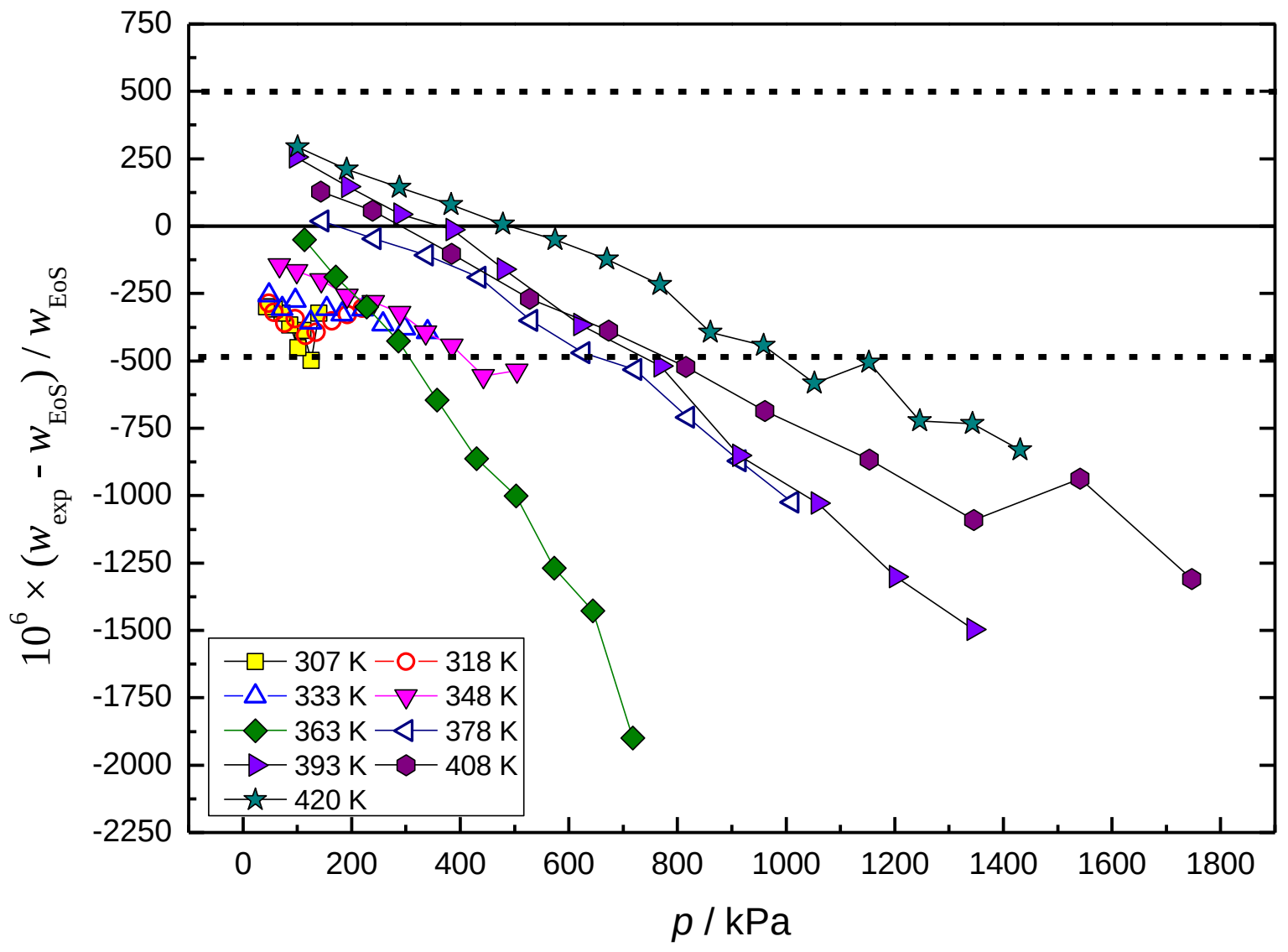

**Fig. 6** – Fractional deviations $(w_{\text{exp}} - w_{\text{EoS}})/w_{\text{EoS}}$ of the speed of sound of R1234ze(Z) measured in this work $w_{\text{exp}}$ from the speed of sound predicted by the EoS of Akasaka et al., 2014; the dashed lines delimit the uncertainty of $w_{\text{EoS}}$.

The more recent formulation by Akasaka and Lemmon (2018, in print), which is based on the larger set of thermodynamic data which became available after 2014, including precise density and sound speed measured in the liquid-phase, virial coefficients, as well as the present data, significantly improves the agreement between the model and this experiment at higher pressures, with an estimated average absolute deviation $\text{AAD} = N^{-1}\sum_{i=1}^{N}\left(\left|w_{\text{exp}} - w_{\text{EoS}}\right|/w_{\text{EoS}}\right)_i$ from the present data AAD = 0.01 %, and an estimated uncertainty of the EoS prediction capability of the speed of sound of 0.02 % in the vapor phase and 0.05 % in the liquid phase.

### 3.2 Ideal gas heat capacities

Squared speed of sound data $w^2(p_i, T)$ obtained at several pressures $p_i$, and corrected at the temperature $T$ of each experimental isotherm, were fitted to a power series expansion of the pressure

$$w^2(p,T) = A_0 + A_1 p + A_2 p^2 + A_3 p^3 + A_4 p^4 + \ldots \ , \tag{5}$$

equivalent to Eq. (1) with $A_0 = \gamma^0 RT/M$, where $\gamma$ is the heat capacity ratio, and the superscript 0 refers to ideal-gas (zero pressure) conditions, with $\gamma^0 = C_p{}^0/C_V{}^0$ and $C_p{}^0/R = \gamma^0/(\gamma^0 - 1)$.

The estimated uncertainties of $w^2(p_i, T)$ are used as weights in their regression to Eq. (5). Figure 7 shows the fractional residuals from this regression for the lowest and highest investigated isotherms, at 307 K and 420 K respectively.

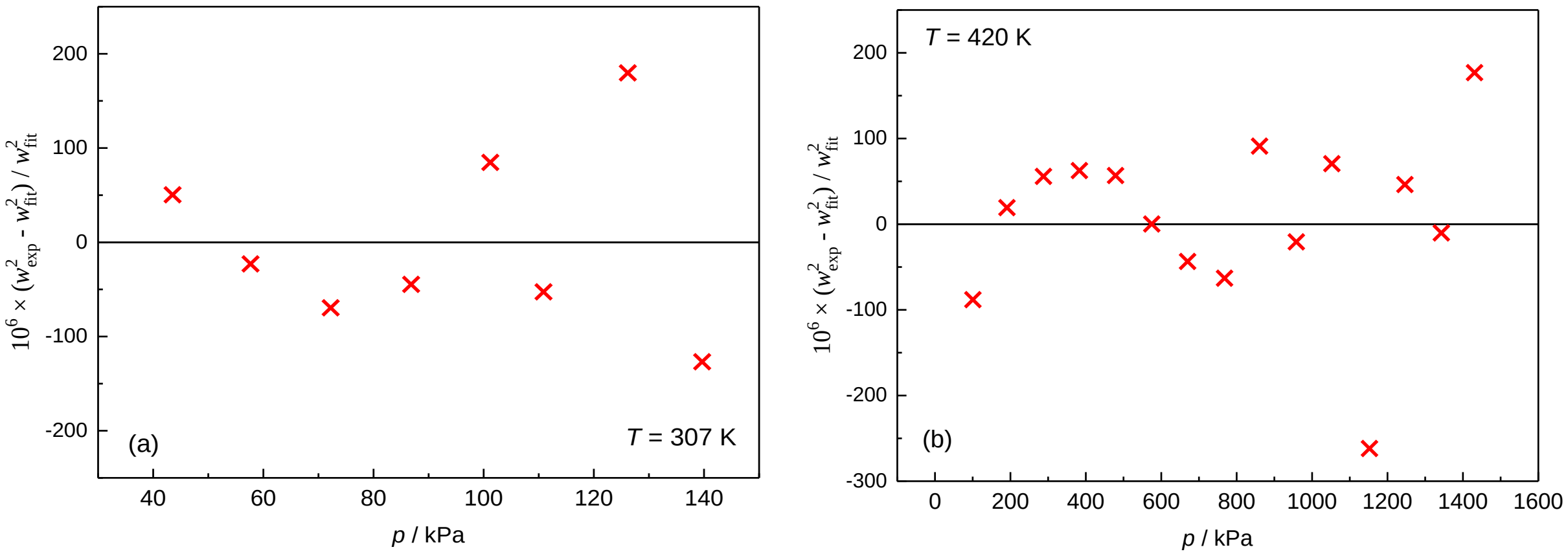

**Fig. 7** − Fractional residuals $\Delta w^2 = (w^2_{fit} - w^2_{exp})/w^2_{exp}$ of squared speed of sound measured in this work from their fit as a function of pressure using Eq. (1) at: (a) $T$ = 307 K, (b) $T$ = 420 K.

The displayed amplitude of the residuals, resulting in the order of a few parts in $10^4$, is of the same order of magnitude of the experimental error. To reduce the residuals to their minimum and satisfactorily represent the data a variable number of terms was needed in the regression with Eq. (5), namely three terms between 307 K and 363 K, four terms between 378 K and 393 K, and five terms for the isotherm at 408 K. Three terms were sufficient for the isotherm at 420 K, where the data collection along the isotherm was prematurely completed and the highest investigated pressure was significantly lower than the saturation pressure at the same temperature.

The ideal-gas isobaric heat capacities $C_p^0$ obtained from these regressions are shown in Fig. 8a and listed in Table 2 together with their relative expanded (coverage factor $k$ = 2) uncertainties resulting from the combination of several uncertainty sources including temperature, molar mass, the standard error of the fitted $A_0$ parameter and, for completeness, the negligible uncertainty[3] of the gas constant $R$.

[3]After May 2019, when the new definition of the kelvin based on an exact value of the Boltzmann constant $k_B$ will be internationally adopted, the uncertainty of $k_B$ and $R$ will be zero.

Table 2 − Ideal-gas heat capacities at constant pressure $C_p^0$ of R1234ze(Z) determined in this work with corresponding relative expanded ($k$ = 2) uncertainties $U_{c,r}(C_p^0/R)$ and fractional deviations from the predicted $(C_p^0/R)_{EoS}$ values from the EoS of Akasaka et al. (2014).

| $T$ / K | $C_p^0/R$ | $10^2 \times U_{c,r}(C_p^0/R)$ | $10^2 \times [C_p^0/R - (C_p^0/R)_{EoS}]/(C_p^0/R)$ |
|---|---|---|---|
| 307.00 | 11.474 | 0.83 | 0.23 |
| 318.00 | 11.834 | 0.36 | 0.56 |
| 333.00 | 12.294 | 0.38 | 0.84 |
| 348.00 | 12.678 | 0.36 | 0.62 |
| 363.00 | 13.058 | 0.74 | 0.51 |
| 378.00 | 13.308 | 0.47 | −0.46 |
| 393.00 | 13.602 | 0.35 | −0.94 |
| 408.00 | 14.012 | 0.52 | −0.48 |
| 420.00 | 14.246 | 0.37 | −0.70 |

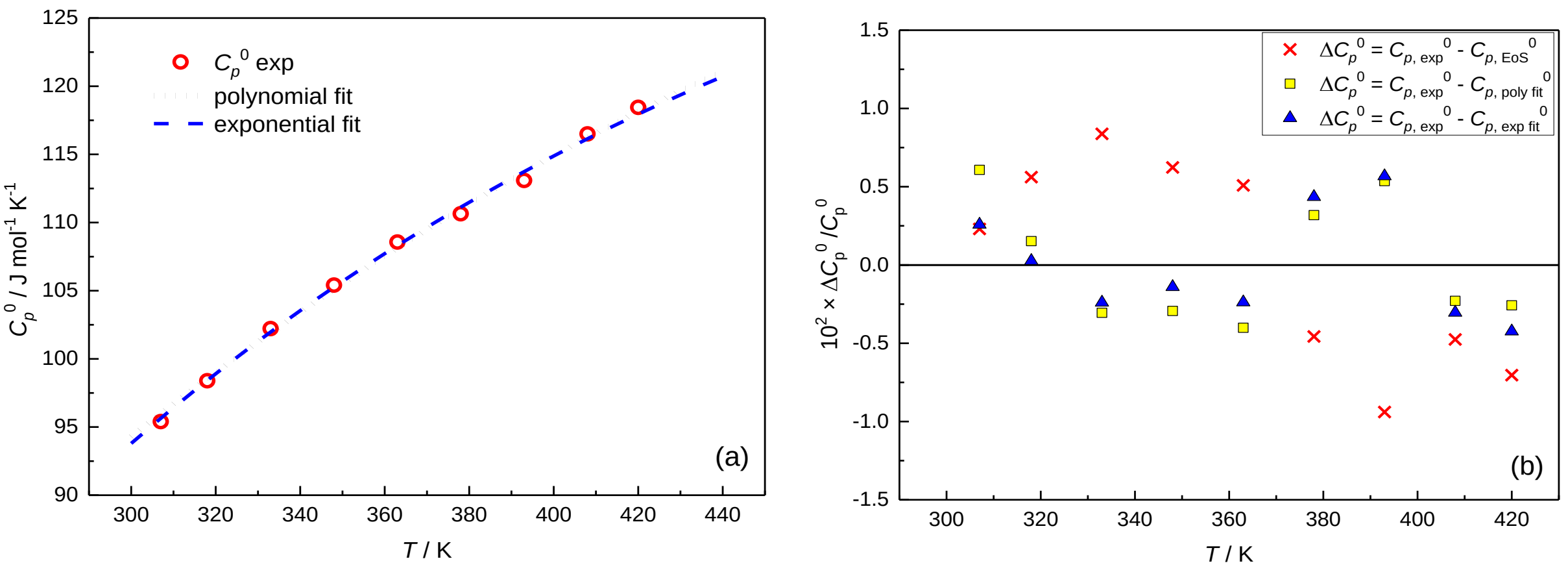

**Fig. 8** − (a) ideal-gas heat capacities at constant pressure $C_p^0$ of R1234ze(Z) as function of temperature. The dashed and dotted lines respectively show the regression curve defined by Eq. (6) and Eq. (7), with the parameters listed in Table 3; (b) crossed symbols ✕ display the relative deviations of the heat capacities determined in this work from those calculated by the EoS of Akasaka et al., 2014; triangles △ show the residuals from a fit using Eq. (6); squared symbols □ show the residuals from a fit using Eq. (7).

To interpolate the complete set of fitted $C_p^0$ as a function of temperature, we tested two alternative functional forms, respectively the non-linear Planck-Einstein function suggested by the molecular theory of the gases:

$$\frac{C_p^{\ 0}}{R} = \nu_0 + \nu_1 \cdot (\nu_2/T)^2 \frac{\exp(\nu_2/T)}{[\exp(\nu_2/T) - 1]^2}, \tag{6}$$

and the second order polynomial expansion:

$$\frac{C_p^{\ 0}}{R} = N_0 + N_1 T + N_2 T^2, \tag{7}$$

by using the estimated uncertainties $u(C_p^0)$ as weights in both cases.

Table 3 − Best fit parameters and standard error of two alternative interpolations of the experimental isobaric heat capacities $C_p^0$ as a function of temperature

| Parameter | Value | Standard Error |
|---|---|---|
| | Non-linear fit with Eq. (6) | |
| $v_0$ | 4 | fixed parameter |
| $v_1$ | 14.80 | 0.15 |
| $v_2$ | − 906.0 K | 9.8 K |
| | Quadratic Fit, Eq. (7) | |
| $N_0$ | − 1.4 | 2.3 |
| $N_1$ | 0.056 K$^{-1}$ | 0.013 K$^{-1}$ |
| $N_2$ | − 4.4·10$^{-5}$ K$^{-2}$ | 1.7 × 10$^{-5}$ K$^{-2}$ |

Figure 8b plots the fractional residuals of these interpolations and Table 3 lists the parameters resulting from these regressions. Both the interpolating functions successfully represent the expected $C_p^0$ values within their experimental uncertainty, with the relative root mean square of the residuals equal to approximately 0.4 % compared to a mean relative expanded ($k$ = 2) uncertainty $U_{c,r}(C_p^0)$ = 0.5 %. Figure 9b also compares our experimental determinations of $C_p^0$ with the predictions of Akasaka et al., (2014), with a resulting root mean squared relative deviation of 0.63%, comparable to the expanded experimental uncertainty reported above. The quality of the agreement is only slightly improved, reducing the same statistical indicator to 0.46 %, by a comparison to the prediction of the more recent EoS of Akasaka and Lemmon (2018 in print), being apparently limited by the uncertainty of our experimental data.

### 3.3 Acoustic virial coefficients

Estimates of the second acoustic virial coefficient $\beta_a$ were obtained, for each isotherm, from the intercept $A_0$ and the linear term $A_1$ of the regression using Eq. (5):

$$\beta_a = RT\left(A_1/A_0\right). \quad (8)$$

These estimates are shown in Fig. 9a and listed in Table 4 with their expanded uncertainties calculated from the combined contribution of the fitted regression parameters, temperature, and their fractional deviations from the EoS of Akasaka et al. (2014). These deviations, with a root mean square of 0.95 %, are found slightly larger than the relative mean expanded ($k$ = 2) uncertainty $U_{c,r}(\beta_a)$ = 0.91 %.

The estimated second acoustic virial coefficients were interpolated as a function of temperature using

$$\beta_a(T) = M_0 + M_1 T^{-1} + M_2 T^{-2} + M_2 T^{-3} \tag{9}$$

with the resulting $M_i$ parameters listed in Table 5. The root mean square of the residuals was found to be 0.59 % in reasonable comparison with the relative mean standard uncertainty reported above.

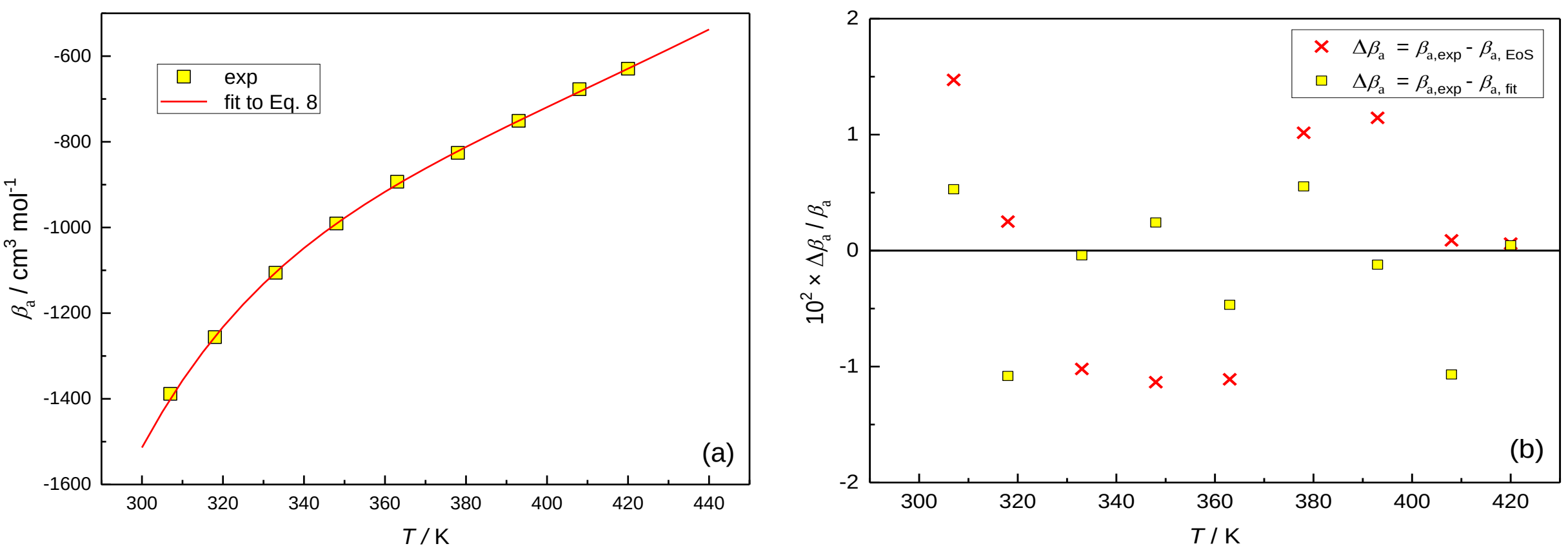

**Fig. 9** – (a): second acoustic virial coefficients $\beta_{a,\,exp}$ of R1234ze(Z) determined in this work. The solid line shows the regression as a function of temperature using Eq. 9; (b) crossed × symbols show the relative deviations $\Delta\beta_a = (\beta_{a,exp} - \beta_{a,EoS})/\beta_{a,EoS}$ from the second acoustic virials predicted by the EoS of Akasaka et al., 2014. Squared □ symbols show the residuals $(\beta_{a,fit} - \beta_{a,exp})/\beta_{a,exp}$ from the interpolating Eq. 9.

Table 4 – Second and third acoustic virial coefficients, $\beta_a$ and $\gamma_a$, corresponding relative expanded ($k$ = 2) uncertainties, relative deviations from the values predicted EoS of Akasaka et al. (2014)

| $T$ / K | $\beta_a$ / cm$^3$ mol$^{-1}$ | $U_{c,r}(\beta_a)$ % | $10^2 \times \frac{(\beta_{a,\,exp} - \beta_{a,\,EoS})}{\beta_{a,\,EoS}}$ | $\gamma_a$ / cm$^6$·mol$^{-2}$ | $U_{c,r}(\gamma_a)$ % | $10^2 \times \frac{(\gamma_{a,\,exp} - \gamma_{a,\,EoS})}{\gamma_{a,\,EoS}}$ |
|---|---|---|---|---|---|---|
| 307.00 | −1388.4 | 2.96 | 1.47 | −7.53·$10^5$ | 76 | -4.9 |
| 318.00 | −1255.9 | 0.70 | 0.25 | −7.19·$10^5$ | 12 | 19 |
| 333.00 | −1105.8 | 0.64 | −1.0 | −6.38·$10^5$ | 8.0 | 49 |
| 348.00 | −990.7 | 0.41 | −1.1 | −4.84·$10^5$ | 4.4 | 58 |
| 363.00 | −893.2 | 0.99 | −1.1 | −3.87·$10^5$ | 8.4 | 74 |
| 378.00 | −825.8 | 0.70 | 1.0 | −1.15·$10^5$ | 32 | −29 |
| 393.00 | −751.1 | 0.34 | 1.2 | −8.9·$10^4$ | 16 | −25 |
| 408.00 | −677.1 | 1.17 | 0.09 | −1.40·$10^5$ | 45 | 63 |
| 420.00 | −629.4 | 0.25 | 0.06 | −8.93·$10^4$ | 4.1 | 36 |

| Table 5 − Parameters defining the interpolation of the second acoustic virial coefficients $\beta_a$ as a function of temperature using Eq. (9) | | |
|---|---|---|
| parameter | value ± standard unc. | unit |
| $M_0$ | (11300 ± 1500) | $cm^3\ mol^{-1}$ |
| $M_1$ | $(-1.25 \pm 0.17)\cdot 10^7$ | $K\ cm^3\ mol^{-1}$ |
| $M_2$ | $(4.51 \pm 0.62)\cdot 10^9$ | $K^2\ cm^3\ mol^{-1}$ |
| $M_3$ | $(-5.75 \pm 0.76)\cdot 10^{11}$ | $K^3\ cm^3\ mol^{-1}$ |

From each regressed $A_2$ term using Eq. (5) the third acoustic virial coefficient $\gamma_a$ at each temperature was obtained as:

$$\gamma_a = (RT)^2 A_2/A_0\,, \tag{10}$$

We remark that for the determination of both $\beta_a$ and $\gamma_a$ the truncation order of regression (5) was the same as that used for the ideal-gas heat capacities.

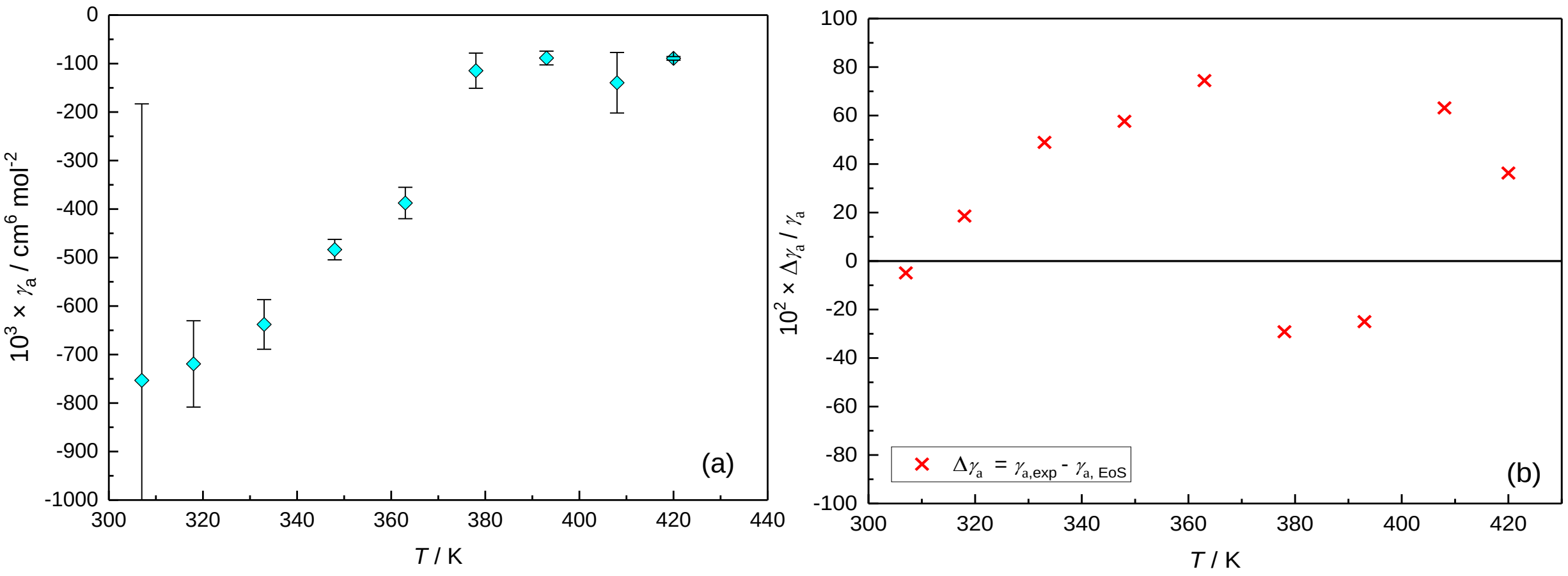

**Fig. 10** − (a) third acoustic virial coefficients $\gamma_a$ with their expanded ($k$ = 2) uncertainties displaying as error bars; (b) relative deviations $\Delta\gamma_a = (\gamma_{a,exp} - \gamma_{a,EoS})/\gamma_{a,EoS}$ from the EoS of Akasaka et al., 2014.

Our estimated third virial coefficients are shown in Fig. 10a and listed in Table 4 with their standard uncertainties, as estimated from the standard error of the regression parameters. Figure 10b shows their relative deviations from the values predicted by Akasaka et al., (2014). The root mean square of the deviations is 45 %, about two times larger than the relative mean expanded ($k$ = 2) uncertainty $U_{c,r}(\gamma_a)$ = 23 %.

**Concluding remarks**

The speed of sound in gaseous R1234ze(Z) has been measured with an accuracy of approximately 0.02 % along nine isotherms in the temperature range 307 K < $T$ < 420 K. From the data, the ideal gas heat capacities were deduced with an estimated expanded ($k$ = 2) uncertainty of approximately 0.5 %. The expanded uncertainty of the second acoustic virial coefficients was approximately 0.9 %, and that of the third acoustic virial coefficients varied between 4 % and 75 % depending on the isotherm. These results were found to be largely consistent with the predictions of the fundamental equation of state of R1234ze(Z) by Akasaka et al. (2014), which was not based on these data, and contributed to a recent, improved equation of state for the same fluid worked out by Akasaka and Lemmon, (2018 in print). Particularly, the agreement of the speed of sound with the prediction of the former equation is in the order of 0.05 %, and is, not surprisingly, improved to 0.01 % by the latter equation, whose stated prediction capability of the speed of sound in gaseous phase is 0.02 %.

We have tried to analyze our acoustic data using a hard-core square well (HCSW) model for the intermolecular potential of R1234ze(Z) in order to determine the lowest order coefficients of a virial density expansion, but contrary to our previous experience, and to several other reports (Goodwin and Moldover, 1991a, 1991b, 1991c, Gillis, 1997), our attempt was unsuccessful yielding to fitted HCSW parameters whose uncertainty was unpractically large. We suspect that the inadequacy of the HCSW model to represent our data, is a consequence of the rather large uncertainty of the acoustic virials used for the interpolation. No attempt with more complex intermolecular potential models was tested.

**Acknowledgements**

We are grateful to Ryo Akasaka, Eric Lemmon and Yuya Kano for sharing their results before publication. We are indebted with Steve Brown for suggesting the idea of this research and to Central Glass co. for providing samples of the fluid.